\documentclass{ws-procs9x6-cpt16}
\begin{document}

\newcommand{\refeq}[1]{(\ref{#1})}
\def\etal {{\it et al.}}

\def\Bq{{\overline B}}
\def\be{\begin{equation}}
\def\ee{\end{equation}}
\def\Bqz{{\overline B}{}^0}
\def\Kqz{{\overline K}{}^0}
\def\cq{{\overline c}}
\def\Aq{{\overline A}}
\def\Re{{\rm Re\,}}
\def\Im{{\rm Im\,}}
\def\Af{A_f}
\def\Afq{{\overline A}{}_f}
\def\re{{\rm e}}
\def\ri{{\rm i}}
\def\bea{\begin{eqnarray}}
\def\eea{\end{eqnarray}}
\def\nn{\nonumber}
\def\babar{$BABAR$~}
\def\babarns{$BABAR$}

\title{Tests of CPT Symmetry in $B^0$-$\Bqz$ Mixing\\ 
and in $B^0 \to c \cq K^0$ Decays}

\author{K.R.\ Schubert}

\address{Johannes Gutenberg-Universit\"at Mainz, 55099 Mainz, Germany}

\author{On behalf of the \babar Collaboration}

\begin{abstract}
Using the time dependences for the decays $\Upsilon(4S)\to B^0\Bqz\to (\ell^\pm X)(c\cq K_{S,L})$, 
we determine the three CPT-sensitive parameters $\Re({\mathsf z})$ 
and $\Im({\mathsf z})$ in $B^0$-$\Bqz$ mixing and $|\Aq/A|$ in $B^0\to c\cq K^0$ decays.
We find 
$\Im ({\mathsf z}) = 0.010 \pm 0.030 \pm 0.013$,
$\Re({\mathsf z}) = -0.065\pm 0.028\pm 0.014$,
and $|\Aq/A|=0.999\pm 0.023\pm 0.017$, in agreement with CPT symmetry. The $\Re({\mathsf z})$ result
provides a limit on one component of Lorentz-symmetry violation.\\
\end{abstract}

\bodymatter

\phantom{}\vskip10pt\noindent
Trying to solve the puzzle of two different $K^+$ mesons with opposite parity, 
Lee and Yang\cite{1956-LeeYang} found in 1956 that there were no convincing tests of P conservation in 
weak-interaction processes. Soon, two experimental groups concurrently proved that P is not only violated in $K^+$ decays, but 
also\cite{1957-Lederman} in the decay chain $\pi^+\to\mu^+\to e^+$ and\cite{1957-Wu} in $\beta$ decays of ${}^{60}$Co. 
CP violation was discovered\cite{1964-CroninFitch} in 1964 in the decays $K^0\to\pi^+\pi^-$ 
at late decay times. In the following years, many authors 
asked if CPT could also be violated in Nature despite its validity in Lorentz-invariant QFT. CP violation implies 
that T or CPT or both are also violated. Bell and
Steinberger\cite{1966-BellSteinberger} proposed in 1965 separate tests of T and CPT using a unitarity relation
with the sum of CP violations in all $K^0$ decay modes. With then all essential inputs measured, 
Bell-Steinberger unitarity resulted in 1970 in\cite{1970-Schubert} $\Re(\epsilon)=(1.7\pm 0.3)\times 10^{-3}$ and 
$\Im(\delta)=(-0.3\pm 0.4)\times 10^{-3}$.
$\Re(\epsilon)$ describes T violation in $K^0$-$\Kqz$ mixing, here established with $\sim 5\sigma$,
and $\delta$ describes CPT violation therein, compatible with zero. In the $B^0$-$\Bqz$ system, large CP violation 
was observed in 2001\cite{2001-BABAR,2001-Belle} in $B^0\to c\cq K^0$ decays, but neither T nor CPT violation has
been observed in $B^0$-$\Bqz$ mixing so far.

Weak-interaction mixing (Standard Model or beyond) in the two-state system $\Psi = \psi_1 B^0+\psi_2 \Bqz$ is described by
the evolution equation
\be 
\ri~\frac{\partial}{\partial t }\left(\begin{array}{c}\psi_1\\ \psi_2\end{array}\right)=
\left[\left(\begin{array}{cc}m_{11}&m_{12}\\ m_{12}^*&m_{22}\end{array}\right)
-\frac{\ri}{2}\left(\begin{array}{cc}\Gamma_{11}&\Gamma_{12}\\
\Gamma_{12}^*&\Gamma_{22}\end{array}\right)\right]
\left(\begin{array}{c}\psi_1\\\psi_2\end{array}\right), \label{Eq-1} 
\ee
with 7 real parameters $m_{11}$, $m_{22}$, $\Gamma_{11}$, $\Gamma_{22}$, $|m_{12}|$, $|\Gamma_{12}|$, and
$\phi(\Gamma_{12}/m_{12})$. Two solutions of Eq.\ \refeq{Eq-1} have an exponential decay law. In lowest
order of ${\mathsf z}$ and $1-|q/p|$, as used throughout this presentation, they are given by
\bea
B_H^0(t) &=& \re^{-\Gamma_H t/2 -\ri m_H t}\left[p(1+{\mathsf z}/2)\,B^0-q(1-{\mathsf z}/2)\,\Bqz\right]/\sqrt{2},\nn\\
B_L^0(t) &=& \re^{-\Gamma_L t/2 -\ri m_L t}\left[p(1-{\mathsf z}/2)\,B^0+q(1+{\mathsf z}/2)\,\Bqz\right]/\sqrt{2},
\label{Eq-2}
\eea
with 7 real observables $m_H$, $\Delta m = m_H-m_L$, $\Gamma_H$, $\Delta\Gamma=\Gamma_H-\Gamma_L$, $|q/p|$, 
$\Re({\mathsf z})$, and $\Im({\mathsf z})$. The 7 observables follow from the 7 parameters, e.g.,
\be
\left|\frac{q}{p}\right|=1-\frac{2\,\Im(\Gamma_{12}/m_{12})}{4+|\Gamma_{12}/m_{12}|^2}~,~~{\mathsf z}=
\frac{(m_{11}-m_{22})-\ri(\Gamma_{11}-\Gamma_{22})/2}{\Delta m-\ri\Delta\Gamma/2}\, .\label{Eq-3}
\ee
In the $K^0$ system, the traditionally used observables are $\Re(\epsilon)=(1-|q/p|)/2$ and $\delta=-{\mathsf z}/2$.
T symmetry requires $|q/p|=1$, and CPT symmetry ${\mathsf z}=0$. From Eqs.\ \refeq{Eq-2}, we obtain the transition rates
as function of the evolution time $t$. With $\Gamma=(\Gamma_H+\Gamma_L)/2$ and $|\Delta\Gamma|\ll\Gamma$, they are
\bea
R(B^0\to B^0) &=& \re^{-\Gamma t}[1+\cos(\Delta m t)-\Re({\mathsf z})\Delta\Gamma t+2\,\Im({\mathsf z})
                  \sin(\Delta m t)]/2\, , \nn\\
R(\Bqz\to \Bqz) &=& \re^{-\Gamma t}[1+\cos(\Delta m t)+\Re({\mathsf z})\Delta\Gamma t-2\,\Im({\mathsf z}) 
                  \sin(\Delta m t)]/2 \, ,\nn\\
R(B^0\to \Bqz) &=& \re^{-\Gamma t}[1-\cos(\Delta m t)]|q/p|^2/2\, ,\nn\\ 
R(\Bqz\to B^0) &=& \re^{-\Gamma t}[1-\cos(\Delta m t)]|p/q|^2/2\, ;\label{Eq-4}
\eea
the first two depend only on ${\mathsf z}$, the last two only on $|q/p|$. Since $\Delta\Gamma$ is unknown,
the first rates determine only $\Im({\mathsf z})$, not $\Re({\mathsf z})$. Transitions into states
decaying into CP eigenstates like $c\cq K_S$, $c\cq K_L$ are also sensitive to $\Re({\mathsf z})$ as shown below. The
world average\cite{2014-PDG} for $|q/p|$ is $1+(0.8\pm 0.8)\times 10^{-3}$. For $\Im({\mathsf z})$, $BABAR$\cite{2006-BABAR}
determined with dileptons $(-14\pm 7\pm 3)\times 10^{-3}$. Using $c\cq K$ decays, $BABAR$\cite{2004-BABAR} found
$\Re({\mathsf z})=(19\pm 48\pm 47)\times 10^{-3}$ in $88\times 10^6\,B\Bq$ events and Belle\cite{2012-Belle}
$(19\pm 37\pm 33)\times 10^{-3}$ in $535\times 10^{6}\, B\Bq$ events. The present analysis from $BABAR$\cite{2016-BABAR}
uses our final data set with $470\times 10^{6}\, B\Bq$ events.

Defining the decay amplitudes $A$ for $B^0\to c \cq K^0$ and $\Aq$ for $\Bqz\to c\cq \Kqz$, with
$\lambda = q\Aq/(pA)$, and assuming (1) $\Delta\Gamma =0$, (2) absence of decays
$B^0\to c\cq \Kqz$ and $\Bqz\to c\cq K^0$, and (3) negligible CP violation in $K^0\Kqz$ mixing, the decay
rates of $B^0$ and $\Bqz$ states into $c\cq K_S$ and $c\cq K_L$ are given by
\be
R_i(t)= N_i\, \re^{-\Gamma t}\, (1+C_i\, \cos\Delta m \,t + S_i\,\sin\Delta m \,t)\, ,\label{Eq-5}
\ee
with,\cite{2016-BABAR} in lowest order of the small quantities ${\mathsf z}$, $|q/p|-1$, and $|\lambda|-1$,
\bea 
C_1(B^0\to c\cq K_L) = +(1-|\lambda|)- \Re(\lambda)\,\Re({\mathsf z})-\Im(\lambda)\,\Im ({\mathsf z})\,,\nn\\
C_2(\Bqz\to c\cq K_L) =-(1-|\lambda|) +\Re(\lambda)\,\Re({\mathsf z})-\Im(\lambda)\,\Im ({\mathsf z})\,,\nn\\
C_3(B^0\to c\cq K_S) = +(1-|\lambda|) +\Re(\lambda)\,\Re({\mathsf z})+\Im(\lambda)\,\Im ({\mathsf z})\,,\nn\\
C_4(\Bqz\to c\cq K_S) =-(1-|\lambda|) -\Re(\lambda)\,\Re({\mathsf z})+\Im(\lambda)\,\Im ({\mathsf z})\,,\nn\\
S_1 = +\Im(\lambda)/|\lambda|-\Re({\mathsf z})\Re(\lambda)\Im(\lambda)+\Im({\mathsf z})[\Re(\lambda)]^2~,\nn\\
S_2 = -\Im(\lambda)/|\lambda|-\Re({\mathsf z})\Re(\lambda)\Im(\lambda)-\Im({\mathsf z})[\Re(\lambda)]^2~,\nn\\
S_3 = -\Im(\lambda)/|\lambda|-\Re({\mathsf z})\Re(\lambda)\Im(\lambda)+\Im({\mathsf z})[\Re(\lambda)]^2~,\nn\\
S_4 = +\Im(\lambda)/|\lambda|-\Re({\mathsf z})\Re(\lambda)\Im(\lambda)-\Im({\mathsf z})[\Re(\lambda)]^2~.\label{Eq-6}
\eea
Neutral $B$ mesons in \babar are produced in the entangled two-particle state $(B^0\Bqz-\Bqz B^0)/\sqrt{2}$ from 
$\Upsilon(4S)$ decays. With a flavor-specific first decay into $\ell^-X$ ($\ell^+X$)%
\footnote{%
In addition to prompt charged leptons 
from inclusive semileptonic decays $\ell^\pm\nu X$,
Ref.\ \refcite{2012-Lees} used charged kaons, 
charged pions from $D^*$ decays 
and high-momentum charged particles 
in the flavor-specific samples $\ell^\pm X$.
}
at time $t_1$, the remaining 
single-particle state is a $B^0$ ($\Bqz$) at this time. Its rate for $c\cq K$ decays at time $t_2=t_1+t$ is given 
by Eqs. \refeq{Eq-5} and \refeq{Eq-6}. 
Decay pairs from the entangled two-particle state respect the two-decay-time formula,\cite{1968-Lipkin} i.e., 
events with the $c\cq K$ decay before the $\ell^\pm X$ decay have the same time dependence
in $t_{c\cq K} - t_{\ell X}$ as events with $\ell^\pm X$ as first decay.
\begin{table}
\tbl{Decay pairs in Ref.\ 15 for the measurement of the decay-time dependences with the coefficients $C_1\cdots C_8$
and $S_1\cdots S_8$.}
{\begin{tabular}
{@{}c c c c c c c c c@{}}
\toprule
$i$ &1 &2 &3 &4 &5 &6 &7 &8\\ \colrule
1st decay &$\ell^- X$ &$\ell^+ X$ &$\ell^- X$ &$\ell^+ X$
          &$c\cq K_L$ &$c\cq K_L$ &$c\cq K_S$ &$c\cq K_S$\\
2nd decay &$c\cq K_L$ &$c\cq K_L$ &$c\cq K_S$ &$c\cq K_S$
          &$\ell^- X$ &$\ell^+ X$ &$\ell^- X$ &$\ell^+ X$\\
\botrule
\end{tabular}}
\label{Tab-1}
\end{table}
With the numeration in Table \ref{Tab-1}, events for $i=$ 5-8 follow Eq.\ \refeq{Eq-5}
with $t=t_{\ell X}-t_{c\cq K}$ and $C_i=C_{i-4}$, $S_i=-S_{i-4}$. The reconstruction of events and the 
determination of the coefficients $C_1 \cdots C_8$, $S_1 \cdots S_8$ are described in Ref.\ \refcite{2012-Lees}.
The obtained values of the 16 coefficients with their uncertainties and correlations 
are used in Ref.\ \refcite{2016-BABAR} for a $\chi^2$ fit of the parameters
$\Im(\lambda)$, $|\lambda|$, $\Im({\mathsf z})$, and $\Re({\mathsf z})$ to the expressions in
Eq.\ \refeq{Eq-6}, leading to the final results
\bea 
\Im(\lambda)=0.689\pm 0.034\pm 0.019~,~~|\lambda|=0.999\pm 0.023\pm 0.017~,\label{Eq-7}\\
\Im({\mathsf z})=0.010\pm 0.030\pm 0.013~,~\Re({\mathsf z})=-0.065\pm0.028\pm 0.014~,\label{Eq-8}
\eea
where the first (second) errors are statistical (systematic). The sign of $\Re({\mathsf z})$
requires the sign of $\Re(\lambda)$. Since $\Im(\lambda)$ and $|\lambda|$ do not fix this sign, additional 
information has to be used%
\footnote{
See Ref.\ \refcite{2016-BABAR} and citations 19-22 therein.
}
for taking $\Re(\lambda)$ negative. The final results are 
independent\cite{2016-BABAR} of the assumption $\Delta\Gamma=0$. Inserting the world 
average\cite{2014-PDG} for $|q/p|$ into the definition of $\lambda$, we obtain
\be
|\Aq/A|=0.999\pm 0.023\pm 0.017\, .\label{Eq-9}
\ee
Under the assumption that $A$ and $\Aq$ have a single weak phase, CPT symmetry 
requires\cite{LeeOehmeYang} $|\Aq/A|=1$.

In conclusion: using $470\times 10^6 B\Bq$ events, \babar finds the results in Eqs.\ \refeq{Eq-8} and \refeq{Eq-9} 
in agreement with
CPT symmetry in $B^0$-$\Bqz$ mixing and in $B\to c\cq K$ decays. The result for $\Re({\mathsf z})$ 
sets a limit on SME coefficients.\cite{Kostelecky, CPT-BABAR}
With $\Delta\Gamma\ll\Delta m$, and averaged over all sidereal times,
\be
\beta^\mu\Delta a_\mu = \gamma\Delta a_0-\beta_z\gamma \Delta a_z = \Re({\mathsf z})\times \Delta m\,,
\ee
where $z$ is defined by the Earth's rotation axis. All $B$ mesons in \babar fly in a very narrow cone 
with\cite{CPT-BABAR}\,$\gamma=1.14$ and $\beta_z\gamma=0.34$, resulting in
\be
\Delta a_0 -0.30\Delta a_z = (-1.9\pm 0.8\pm 0.4)\times 10^{-14}\,\rm GeV.
\ee

\end{document}